
\def\to{\rightarrow}
\def\ds{\displaystyle}
\input phyzzx
\sequentialequations
\PHYSREV

\date{ }
\titlepage
\title{ Low-energy effects of Charged Higgs with general Yukawa couplings }
\author{ J.L. Diaz Cruz$^{a,b}$, J.J. Godina Nava $^a$, G. Lopez Castro $^a$ }
\address{ a) Depto. de Fisica, CINVESTAV-IPN, Ap. Postal 14-740,
07000 M\'exico D.F. \break
b) Instituto de Fisica,
U.A. de Puebla, Puebla, Pue., M\'exico}

\abstract
We study a model with two Higgs doublets where FCNC are allowed at tree-level.
In this model, the interactions of charged Higgs with fermions
($H^\pm f \bar{f'}$),
include a term that is not proportional to the fermion masses,
which we constraint using the following low-energy processes:
i) tau decays
($\tau \to \nu_{\tau} e\nu_e, \nu_{\tau}\mu\nu_{\mu}, \nu_{\tau}\pi$),
ii) leptonic decays of pseudoscalar mesons ($\pi, K \to \ell\nu_{\ell}$) and
iii) semileptonic b-decays.
With these constraints it is possible to make predictions;
we illustrate this by  presenting the rates for
the (FCNC) decay $c\to u+\gamma$,
the (second class-current) decay $\tau\to \nu_{\tau}+ \eta\pi$,
and also the theoretical value of the neutron life-time.

\endpage


\FRONTPAGE

\Ref\smth{S. Weinberg, Phys. Rev. Lett. 19 (1967) 1264; A. Salam in
Elementary Particle Theory, ed. N. Southolm (Almquist and
Wiksell, Stockholm, 1969), p. 367; S.L. Glashow, Nucl. Phys. 22
(1961) 579.}
\Ref\topl{F. Abe et al., CDF Collaboration,
Phys. Rev. Lett. 73 (1994) 225.}
\Ref\hlep{ G. Coignet, Proceed. of the XVI Lepton Photon Symposium, Eds.
P.Drell and D. Rubin, Cornell Univ. August 10-15 (1993).}
\Ref\kmm{ M. Kobayashi and T. Maskawa, Prog. Theor. Phys. 49 (1973) 652.}
\Ref\cpsb{ G. Branco and M. Rebelo, Phys. Lett. 160B (1985) 117. }
\Ref\gim{ S. Glashow, J. Iliopoulus and L. Maiani,
Phys. Rev. D2 (1970) 1285.}
\Ref\pdat{ Particle Data Book, Phys. Rev. D45 part II (1992).}
\Ref\tacp{ G. Eilam et al., Phys. Rev. Lett. 67 (1991) 1979.}
\Ref\tcga{ J.L. Diaz-Cruz et al., Phys. Rev. D41 (1990) 891;
      G.Eilam et al., Phys. Rev. D44 (1991) 1473. }
\Ref\txhx{J.L. Diaz-Cruz and G. Lopez Castro, Phys, Lett, B301 (1993) 405;
M. Luke and M. Savage, Phys. Lett. B307 (1993) 387.}
\Ref\thdm{ H. Georgi, Hadronic Jour. 1 (1978) 155.;
 H. Haber, G.L. Kane, T. Sterling, Nucl. Phys. B161 (1979) 493;
 S. Bertolini, Nucl. Phys. B272 (1986) 77;
J.F. Gunion, H.E. Haber, Nucl. Phys. B272 (1986) 1 .}
\Ref\hunt{For complete references see: J. Gunion et al., "The Higgs Hunter
Guide",  Addison-Wesley.}
\Ref\syu{ M. Sher and Y. Yuan, Phys. Rev. D44 (1991) 1461.}
\Ref\clin{For an elementary presentation see: D. Cline, Scientific Am, Sept.
1994.}
\Ref\hou{ W-S. Hou, Preprint PSI-PR-91-34 (1991);
T.P. Cheng and M. Sher, Phys. Rev. D35 (1987) 3448; A. Antaramian L.J. Hall and
A Rasin, Phys. Rev. Lett. 69 (1992) 1871; L.J. Hall and S. Weinberg, Phys. Rev.
D48 (1993) 979.}
\Ref\pols{ P. Krawczyk and S. Pokorski, Nucl. Phys. B364 (1991) 10.}
\Ref\taexp{ A. Schwarz, in Ref. 3}
\Ref\tatu{W. Marciano and A. Sirlin, Phys. Rev. Lett. 61 (1988) 1815.}
\Ref\mars{W. Marciano and A. Sirlin, Phys. Rev. Lett. 71 (1993) 3629; D.A.
Bryman, Comments on Nuc. and Part. Phys. 21 (1993) 101.}
\Ref\fink{R. Decker, M. Finkemaier, Karlsruhe preprint TTP-53-25 (1993).}
\Ref\bigi{I. Bigi, B. Blok, M.A. Shifman and A. Vainshtein, CERN-TH.7082/93;
UND-HEP-93-BIG06; UMN-TH-1225/93; TPI-MINN-93/53-T;
Technion-Ph-93-40; October 1993.}
\Ref\gris{B. Grinstein, N. Isgur, M.B. Wise, Phys. Rev. Lett. 56 (1986) 298; M.
Wirbel B. Stech, M. Bauer, Z. Phys. C29 (1985)627; J.G. K\"orner and G.A.
Schuler, Z. Phys. C46 (1990) 93.}
\Ref\venu{V. Luth, in Ref. 3.}
\Ref\cleo{CLEO Collaboration, A. Bean et. al., Phys. Rev. Lett. 70 (1993)
2681.}
\Ref\alep{ALEPH Collaboration, contributed paper XXVI Intern. Conf. on High
Energy Physics (Dallas, TX, August 1992).}
\Ref\isi{ G. Isidori, Phys. Lett. B298 (1993) 409. }
\Ref\egli{S. Egli, C. Grab, F. Ould-Saada, H. Simma, D. Wyler, Preprint
ETHZ-IMPPR/92-1.}
\Ref\wei{ Wei-Shu Hou and R.S. Willey, Nucl. Phys. B326 (1989) 54; N. G.
Deshpande and G. Eilam, Phys, Rev. D26 (1982) 2463.}
\Ref\taus{Workshop on the Tau-Charm-Factory, Marbella, Spain, June (1993).}
\Ref\tuso{ M. Artuso et. al., CLEO Collaboration, Phys. Rev. Lett. 69 (1992)
3278.}
\Ref\ymau{Y. Meurice, Phys. Rev. D36 (1987) 2780;
C.A. Dom\'{\i}nguez, Phys. Rev. D20 (1979) 802;
Y. Meurice, Mod. Phys. Lett. A2 (1987) 699; Phys. Rev. D36 (1987) 2780;
A. Bramon, S. Narison and A. Pich, Phys. Lett. B196 (1987) 543;
 A. Pich, Phys. Lett. B196 (1987) 561;
E. Berger and H. J. Lipkin, Phys. Lett. B189 (1987) 226;
J.L. Diaz Cruz and G. L\'opez Castro, Modern Phys. Lett. A6 (1991) 1605.}
\Ref\yan{Y. Meurice, Phys. Rev. D36 (1987) 2780.}
\Ref\cast{ See for example: D. Thompson, J. Phys. G: Nucl. Part. Phys. 16
(1990) 1423.}
\Ref\habe{H. Haber, G. L. Kane, T. Sterling, Nucl. Phys, B161 (1979) 493.}
\Ref\wilk{D. Wilkinson, Nucl. Phys. A377 (1982) 474.}

\endpage

\chapter{\bf Introducci\'on }

The standard model (SM) of electroweak interactions [\smth] has been
quite succesfull in confronting the experiments,
although it requires the discovery of the top quark and Higgs boson
for its confirmation.
The existence of the top is widely accepted,
and there are already signs of its presence at FNAL,
with a mass of 174 GeV [\topl]. On the other hand,
LEP has established a lower bound on the Higgs boson mass of
63.5 GeV [\hlep].

 Because of several puzzles that the SM can not
explain, e.g. the number of parameters, the fermion family and mass patterns,
the hierarchy and naturalness problems etc., many alternatives or
extensions to the SM have been proposed.
In some of these extensions there are new sources of CP-violation,
besides the SM one (which comes from the KM mixing matrix [\kmm]), for instance
spontaneous CP-violation [\cpsb].
Some extensions contain also flavour changing neutral currents (FCNC) at
tree-level,
which are absent in the SM [\gim].

In fact, CP-violation and FCNC decays have been observed for the down-type
quark sector only [\pdat]. For up-type quarks, the SM predicts
that CP and FCNC phenomena will occur at very small rates;
 for instance, the SM prediction for the
CP asymmetry in top decays is very small,
about $10^{-11}$ for the decay $t\to b\bar b c$ [\tacp].
Similarly, the SM result for the FCNC decay
$t\to c+\gamma$ gives a very small branching ratio (BR), about
$ 10^{-12}$ [\tcga], which will not be detectable at future
experiments. However, in extensions of the SM with an enlarged
Higgs sector, the BR for $t\to c+\gamma$ can reach a value of
$10^{-6}$, and the CP asymmetry for the decay $t\to b+\tau^+ \nu$
can be about 6 orders of magnitude larger than the SM value, namely
$O(10^{-5})$ [\txhx], which could be tested at future colliders.

In models with two Higgs doublets ($\Phi_1$ and $\Phi_2$) [\thdm], it was usual
to avoid FCNC mediated by neutral Higgs bosons,
by imposing a discrete symmetry
$\Phi_1 \to \Phi_1,\ \ \ \Phi_2 \to -\Phi_2 $, which could be broken
only softly (i.e. by terms of dimension two). The absence of FCNC
was imposed in these models because of the belief that
in order to satisfy the current experimental limits,
the Higgs bosons have to be very heavy, which in turn
would lead to violations of unitarity [\hunt].
However, as it was noticed in [\syu],
FCNC can also be suppressed with relatively
light Higgs masses ($m_h< O(1\ TeV$)), provided that the
couplings $hq_i\bar q_j $ behaves like $(m_im_j)^{1/2}$
instead of $m_i^2$, as it was assumed before.

In this paper we are interested in
evaluating the constraints that can be imposed on the parameters
of a two-Higgs doublet model that allows FCNC at tree-level,
particularly in the charged Higgs parameters, using present low-energy data.
Once these constraints are known, one can use them to make quantitative
predictions for new phenomena, which will be the signature of the new physics
incorporated in the model [\clin].

The neutral Higgs sector of two-Higgs doublet models with FCNC, has been
already studied in Refs. [\syu,\hou], focusing on both FCNC and CP
phenomena.
It has been also shown in [\hou], that the type of Yukawa couplings
needed to supress FCNC, can be obtained in models where the fermion masses are
derived using some ansatz for the mass matrices, e.g. the Fritzch type.

We shall follow a similar approach as in [\syu,\pols],
where the focus is in constraining the couplings that would
signal the new physics, rather than obtaining limits only
on the Higgs masses, as has been usual in most studies. We follow
this approach because our motivation is to use the properties of the charged
Higgs, as a tool to test the models that attempt to explain the fermionic mass
pattern.

The organization of this paper is as follows.
In section 2 we explain some details of the model.
Section 3 contains our main results, namely the analysis
 of several low-energy
processes used to constraint
the interactions of the charged Higgs with fermions.
The data used in the work has been selected mostly to make a first
evaluation of the relevant parameters of the model,
rather than to do an extensive
analysis of the parameter space.
These constraints are used in section 4 to predict
the rates for: the FCNC decay $c\to u+\gamma$,
and the (second class current) decay $\tau^-\to \nu_{\tau}+ \eta+\pi^-$.
We consider also the apparent discrepancy between the
SM prediction and the measured neutron life-time, to see
if it could be resolved within our model.
We have choosen these processes because they are sensitive
to the appearence of a charged scalar, although there may be
other ones where the predictions could be  even more sensitive.
Finally, our conclusions are
presented in sect. 6.

\chapter{\bf The Model}

The model includes two Higgs doublets, $\Phi_1$ and $\Phi_2$,
of equal hypercharges (Y=1), such that after spontaneous symmetry
breaking it contains five Higgs bosons. The most general Yukawa interaction
lagrangian can be written as follows
\foot{A similar piece of the lagrangian should be understood for
the Yukawa couplings of the leptons.}:
$$L_Y= \overline{\Psi}^0_{Li} ( F_{ij}\tilde{\Phi_1}+
            \xi F'_{ij}\tilde{\Phi_2} ) U^0_{Rj} +
        \overline{\Psi}^0_{Li} ( G_{ij}{\Phi_2}+
            \xi G'_{ij}{\Phi_1} ) D^0_{Rj} + h.c. \eqno\eq $$
where $\Psi^0_{Li}= ( U^0_i, D^0_i)_L$ denote the $SU(2)$ quark doublet, and
$U^0_{Rj},\ D^0_{Rj}$ are the quark singlets (the superscript 0 is used for
weak eigenstates, whereas $i,j$ denote generation number).
$F, F', G$ and $G'$ are dimensionless $3\times 3$ matrices
characterizing the Yukawa couplings and $\xi$ parametrizes the
discrete symmetry breaking.

A $U(1)_{em}-$invariant VEV is given by
$<\Phi_1>^T= (0, {\ds v_1\over \ds \sqrt{2} } )$,
$<\Phi_2>^T= (0, {\ds v_2e^{i\delta}\over \ds \sqrt{2}})$
where the phase $\delta$ signals the violation of CP in the Higgs sector.
The scalar spectrum consists of a charged Higgs pair ($H^\pm$)
and three neutral Higgs bosons ($A^0,h^0,H^0$).
After SSB, we obtain from Eq.(2.1)
the following expressions for the
fermion mass matrices,
$$ M^0_u = (F+e^{-i \delta} \xi {v_2\over v_1} F')
{v_1\over \sqrt{2}} \eqno\eq $$
$$ M^0_d = (G+e^{-i \delta} \xi {v_1\over v_2} G')
{v_2\over \sqrt{2}}. \eqno\eq $$
In order to diagonalize these mass matrices we shall work in a
basis where $M^0_u$ is diagonal ($M^0_u=M_u$).
Then the mass matrix for down-type quarks can be diagonalized
through the following transformations,
$$M_d=V^+_L M^0_dV_R={v_2\over \sqrt{2}}
V^+_L[ G + e^{-i\delta}\xi {v_1\over v_2}G']
                           V_R \eqno\eq $$
$V_L$ can be identified as the KM mixing matrix.
Then, the interaction between the quarks (mass eigenstates) and
the charged Higgs boson is given
by the following expresion,
$$ \eqalign{ L_{q_i\bar q_jH^\pm}= & {g\over \sqrt{2}m_W}H^+ \overline{U}
    [\cot\beta V^+_L M_d R  + \tan\beta V_{L}M_u L+  \cr
   & \xi e^{-i\delta_1} M_1\Gamma L+ \xi e^{-i\delta_2}M_2\Gamma'R ]
      D + h.c. ,\cr } \eqno\eq $$
where $\Gamma=F'V_L$, $\Gamma'=G'V_L$ and (L, R) denote the helicity projection
operators.
The phases $\delta_1, \delta_2$ are given by:
$$\tan\delta_1=-{\sin\delta\over (\cos\delta+\cot\beta)}, $$
$$\tan\delta_2=-{\sin\delta\over (\cos\delta+\tan\beta)}. $$
The parameters $M_{1,2}$ are given by:
$M_1=\sqrt{2}m_W s_1/g$ and
$M_2= \sqrt{2}m_W \cot\beta s_1/g$, \nextline with $\tan \beta\equiv v_2/v_1$,
$s_1=[sec^2\beta-4\sin^2 \beta \sin^2\delta/2 ]^{1/2} $,
 and $g$ denotes
the SU(2) coupling constant.

A similar expression is obtained for leptons. In this paper we assume that
neutrinos are massless, then one can choose a basis where the coupling between
the $W^{\pm}$ and leptons is diagonal. On the other hand, because of the terms
that produce FCNC, the
 couplings $H^+\ell\nu_{\ell}$ are not diagonal in general.
However, in the present work we shall neglect the effects of the non-diagonal
terms in the $H^+\ell_i\nu_j$ vertex.

\chapter{\bf Bounds on the Yukawa couplings of the charged Higgs}

In this section we shall study the limits that can be imposed on the
charged Higgs-fermions couplings from low-energy data. The focus will be on the
quantities that will be used later for the predictions of the
model. We shall attempt to obtain limits only
for some sets of typical values of parameters,
rather than doing an extensive analysis of the complete
regions of parameter space. We believe that this will illustrate
the most important features of our model.

In the following analysis, we shall neglect in the
vertices $H^{\pm} f_i \overline {f_j}$ the contributions proportional
to the masses of s- and b-quarks, with respect to c- and t- masses.
{}From an analysis of Yukawa interactions using
some ansatz for the mass matrices [\hou], it was found that
 $\Gamma'$ (in Eq.(2.5)) is neglibible with respect to $\Gamma$,
because of the hierarchy of quark masses and mixing angles.
Thus, with this approximation  the vertex
$H^{\pm}f_i \overline {f_j}$ for light fermions reduces to the form:
$$ V_{H^{\pm}f_i \bar {f_j}}= {g m_H\over 2m_W}e^{i\delta_1}
\lambda_{ij}(1-\gamma_5),  \eqno\eq $$
where $\lambda_{ij}=\xi M_1\Gamma_{ij}/m_H$.
The dimensionless
factors $\lambda_{ij}$ appear in a natural way in our calculations
of the following sections, because we are considering weak decays of light
particles ($m_f<< m_{H^\pm}$) and then our effective lagrangian
corresponds to a four-fermion local interaction. Our results are expressed as
bounds for $\lambda_{ij}$ which are summarized in table 1.

\section{Constraints from leptonic tau decays}

The recent experimental results on the mass and leptonic branching ratios
of the tau have reached such precision [\taexp], that they allow now
to derive bounds on new physics. These measurements will be
used to obtain bounds on the parameters $\lambda_{e\nu}$, $\lambda_{\mu\nu}$,
$\lambda_{\tau\nu}$ defined in Eq.(3.1).

The width for the decay $\tau \to \nu_{\tau}+ l+\nu_l$,
including $W^\pm$ and $H^\pm$ contributions can be written as follows,
$$ \Gamma= {G^2_Fm^5_{\tau}\over 192\pi^3} f(z_l)(1+ h_{RC})
[1+{\lambda^2_{\tau\nu_{\tau}}\lambda^2_{\ell\nu_{\ell}}\over 4} ], \eqno\eq $$
where $z_l= {\ds m^2_l\over \ds m^2_{\tau}}$,
$f(z)=1-8z+8z^3-z^4-12z^2logz$. The effect of radiative corrections
\foot{In the calculations of the present and forthcoming
sections for SM allowed processes, we shall use the
approximation that radiative corrections are the same for
the amplitudes mediated by $W^{\pm}$ and $H^\pm$, since this accounts to
neglect small terms of order $O(\alpha\lambda_{ij}\lambda_{l\nu})$.}
is included in the function $h_{RC}$, which has been evaluated in [\tatu].

The previous decays are clean predictions of the SM if one uses the value
of  $G_F=1.16637(2)\times 10^{-5}$ GeV$^{-2}$, as obtained from muon decay.
In order to bound the charged Higgs contribution, we shall
require it to lay below the experimental uncertainty.
\foot{This procedure will be used also in the forthcoming sections, unless
otherwise specified.}
Using the data: $BR(\tau \to e\nu_e\nu_{\tau})=0.1789\pm 0.0014$,
$BR(\tau\to \mu\nu_{\mu}\nu_{\tau})=0.1734\pm0.0016$ [\taexp], we obtain
the bounds that appear in the first two entries of table 1, namely:
$$ \lambda_{\tau\nu_{\tau}}\lambda_{e\nu_e} < 0.22 \eqno\eq $$
$$ \lambda_{\tau\nu_{\tau}}\lambda_{\mu\nu_{\mu}} < 0.23 \eqno\eq $$
For completeness, the bound obtained from muon decay is also given in table 1
\foot{If one considers non-diagonal terms in the vertex $H^+\ell_i\nu_j$,
then the previous bound would translate into a bound for
$\sum_{i,j} \lambda_{\tau \nu_i} \lambda_{\ell \nu_j}$.}.

As it will be explained in the following sections,
one does not have to know the bounds separately for each of the
$\lambda_{ij}$'s since only the previous combination
will appear in the quantities of interest in the present paper.

\section{Constraints from leptonic decays of pseudoscalar mesons}

The leptonic decays of charged pseudoscalar mesons $P^\pm \to l^\pm+\nu_l$,
$(P= \pi, K)$, can be used to obtain bounds for
the products: $\lambda_{ud}$$\lambda_{l\nu_{l}}$,
$\lambda_{us}$$\lambda_{l\nu_{l}}$.
According to [\mars], the SM result for the decay width of the meson P, can be
written as follows:
$$ \Gamma_{SM}= {G^2_F|V_{uj}|^2 f^2_P m_P m^2_l \over 8\pi}
\big[1-{m^2_l\over m^2_P}\big]^2 (1+ h_{RC}), \eqno\eq $$
where $f_P$ is the decay constant of the pseudoscalar meson $P$,
which is defined from:
$<0|\bar q_j(x)\gamma_\mu\gamma_5 u|P>= if_Pp_{\mu}e^{-ip.x}$
($ q_j=d, s$ for $\pi^+$ and $K^+$ mesons, respectively);
the function $h_{RC}$ includes the effect of radiative corrections [18].

The tree-level contribution of the charged Higgs to this decay width
is obtained by adding the graph where the W boson is substituted
by the charged Higgs. Because of parity-invariance,
only the pseudoscalar part of the Higgs-fermion interaction
contributes to the amplitude, which is then given by:
$$ M= {G_F\over \sqrt{2}}
     \big[V_{ju}<0| \bar q_j(x)\gamma_{\mu}\gamma_5 u|P>l_{\mu}
+\lambda_{ju}\lambda_{\ell\nu}<0|\bar q_j(x)\gamma_5 u|P> l_S\big] \eqno\eq $$
where the leptonic currents $l_{\mu}$ and $l_S$ are given by:
$ l_{\mu}= \bar l\gamma_\mu(1-\gamma_5)\nu_l $, \nextline
$ l_S= \bar l(1-\gamma_5) \nu_l$, respectively.

The hadronic matrix element of the pseudoscalar current can be obtained
from the axial one by using the relation:
$$<0|\bar{q_j}\gamma_5 u|P>\equiv {if_Pm^2_P\over (m_j+m_u)}=A_1V_{ju}
. \eqno\eq $$
where $A_1={\ds 2 m^2_P \over \ds {V_{ju} m_{\ell}(m_j+m_u)}}.$

Then, the total width is given by,
$$ \Gamma= \Gamma_{SM} (1+2A_1 \lambda_{ju}\lambda_{\ell\nu}
       + A^2_1 \lambda^2_{ju}\lambda^2_{\ell\nu} ). \eqno\eq $$

It has become convenient in tests of $e-\mu$ universality, to evaluate  the
ratio:
$$R_{e/\mu}(P)={\Gamma(P\to e\nu_e(\gamma))\over \Gamma(P\to
\mu\nu_{\mu}(\gamma))},$$
which is independent of $f_P$. By using the extreme values allowed by
one standard deviation in the ratios:\nextline
$R^{exp}_{e/\mu}(\pi)/R^{SM}_{e/\mu}(\pi)=0.9966\pm 0.0031$ [\mars], and
$R^{exp}_{e/\mu}(K)/R^{SM}_{e/\mu}(K)=0.965\pm 0.043$ [\pdat], we obtain the
following bounds:
$$ \mid\lambda_{ud}(\lambda_{e\nu}-{m_e\over m_{\mu}}\lambda_{\mu\nu})\mid<6.44
\times 10^{-7},$$
$$\mid \lambda_{us}(\lambda_{e\nu}-{m_e\over m_{\mu}}\lambda_{\mu\nu})\mid<1.84
\times 10^{-6}. $$

\section{Constraints from the decay $\tau\to \nu_{\tau}+\pi$}

We will use this decay to get a bound on the product
$\lambda_{ud}\lambda_{\tau\nu}$. The contribution from the charged
Higgs to the decay amplitude can be included along similar lines of the
previous sections. In order to use the radiative corrections
from Ref. [\fink,\tatu], we will work out the following ratio:
$${\Gamma(\tau \to \nu_{\tau}+\pi) \over \Gamma(\tau\to e\nu_e\nu_{\tau})}=
{\Gamma^{SM}(\tau\to \nu_{\tau}+\pi)[ 1+\delta_{\tau\pi}] \over
 \Gamma^{SM}(\tau\to e\nu_e\nu_{\tau})[1+\delta_{\tau e}] } \eqno\eq $$
where the SM contribution has been factored out, and
the terms that include the charged Higgs contributions
are given by:
$$ \delta_{\tau\pi}= {4m^2_{\pi} \lambda_{ud}\lambda_{\tau\nu}\over
                         m_{\tau} V_{ud}(m_d+m_u)} , \eqno\eq $$
and $ \delta_{\tau e}= \lambda^2_{\tau\nu} \lambda^2_{e\nu}/4,$
which has been bounded in section 3.1.

In order to bound $\delta_{\tau\pi}$, we proceed as follows:
for the left-hand side of Eq.(3.9) we use the experimental data [\pdat,\taexp],
$$ {\Gamma^{exp}(\tau \to \nu_{\tau}+\pi) \over
 \Gamma^{exp}(\tau \to e \nu_e\nu_{\tau})} = 0.648\pm0.023.$$
For the SM part we use the result of Ref. [\fink], which includes $O(\alpha)$
radiative corrections, namely

$${\Gamma^{SM}(\tau \to \nu_{\tau}+\pi) \over
\Gamma^{SM}(\tau\to e\nu_e\nu_{\tau})}= 0.619\pm0.001.$$

Then including the bound (3.3) for $\delta_{\tau e}$,
we find $\delta_{\tau\pi}= 0.035\pm 0.039 $, which can be translated
into a limit for the product of couplings:

$$ \lambda_{ud}\lambda_{\tau\nu}< 2.6 \times 10^{-2}, \eqno\eq $$

Let us comment that this particular combination of couplings will appear
in the evaluation of the second-class decay of the tau (section 4.2).

\section{Constraints from the decay $b\to u+\ell\nu_{\ell}$ }

  The measurements of the decay rates of heavy quarks
can also be used to extract information on the couplings
$\lambda_{Qq}$. In particular, there are measurements of the
inclusive decay $B\to X \ell\nu_{\ell}$, which
allow to put bounds on the coupling $\lambda_{bu}$.

 If we consider only the cases when $l=e,\mu$, then the decay
width for the semileptonic b decay including $H^\pm$ contributions, is written
as follows:

$$\Gamma(b \to u+ \ell+\nu_{\ell})=\Gamma_0 f(z_l) \eta
|V_{ub}|^2[1+{\lambda^2_{\tau\nu} \lambda^2_{e\nu}\over 4 |V_{ub}|^2}] \eqno\eq
$$
where $z_l=m^2_{\ell}/m^2_b$, $\Gamma_0= G^2_Fm^5_b/192\pi^3$, and
the phase space factor $f(z)$ is defined in Eq.(3.2).

For light leptons ($\ell=e,\mu$), one can set $f(z_{\ell})\simeq1$.
The factor $\eta$ in Eq.(3.12) includes effects from perturbative QCD
corrections, and has been estimated
to be $\simeq0.86$ [\bigi].

In order to extract the bound on the factor $\lambda_{bu}\lambda_{e\nu}$,
we use the theoretical predictions for the ratio
$R_{B/b}={\ds \Gamma(B\to \rho+e\nu_e)\over \ds \Gamma(b\to u+e\nu_e)}
\simeq 3.5-14 \% $ [\gris], which can be combined with the measured values of
the life-time $\tau_B=(1.49\pm0.04)\times 10^{-12}$ sec [\venu] and $BR(B\to
\rho+l\nu_l)< (1.6-2.7)\times 10^{-4}$ [\cleo],
as follows:
$$\Gamma(b\to u+\ell\nu_{\ell})={ BR(B\to \rho+\ell\nu_{\ell})
                  \over \tau_B R_{B/b}  } . \eqno\eq $$
Since $|V_{ub}|$ is poorly known, and in fact is expected to be obtained
from the previous decay, we shall neglect the SM contribution in order
to estimate a bound on $\lambda_{bu}$. We obtain (for $m_b=4.8~GeV$):
$$\lambda_{bu}\lambda_{e\nu} < 1.43 \times 10^{-2}, $$
which appears in the summary table 1.

\section{Constraints from the decay $b\to c+\tau\nu_{\tau}$.}

To extract the parameter $\lambda_{bc}$, which corresponds
to the remmaining  $H^{\pm}f_i\overline f_j$ couplings needed
to make our predictions,
we shall use the decay $b\to c+\tau \nu_{\tau}$.

The BR for this decay has been meassured recently [\alep],
($BR(b\to c+\tau \nu_{\tau}) = (4.2^{+0.72}_{-0.68}\pm0.46)\times 10^{-2} $), a
result that
is in fact significantly above the SM prediction
\foot{The decay width for
$\Gamma(b\to c+\ell\nu_{\ell}), \ell= e,\mu$, can not be used because its
value is used to fix $V_{bc}$.}
 $(BR\simeq 2.5 \%)$.

The amplitude for the decay width, including $W^\pm$ and $H^\pm$ contributions,
is evaluated along the same lines as before. However,
in order to evaluate the 3-body phase space for this case,
with two masses in  the final state that can not be neglected,
we choose to perform a numerical integration.
Thus, the decay width can be written as follows,
$$ \Gamma(b\to c \tau\nu_{\tau})=\Gamma_{SM}(1+{
\lambda^2_{bc}\lambda^2_{\tau\nu}\over {4 \mid V_{bc}\mid ^2}}), \eqno\eq $$
where $\Gamma_{SM}$ has the following form,

$$\Gamma_{SM}={ G^2_F  m^5_b\mid V_{bc}\mid ^2\over { 192 \pi^3}}
f(\xi_{\tau},{\xi_c})\left[1-{2\alpha_s(m_b)\over 3\pi}(\pi^2-{25\over
4})\right] \eqno\eq$$
where
$$\eqalign{ f(\xi_{\tau},{\xi_c})=&12{\ds
\int^{1+\xi_c-\xi_{\tau}}_{2\sqrt{\xi_c}}dx\int^{y^+}_{y^{-}}dy W(x,y)}\cr
=&12{\ds \int^{1+\xi_c-\xi_{\tau}}_{2\sqrt{\xi_c}}dx\int^{y^+}_{y^{-}}dy
H(x,y)}}$$
with $\xi_i={\ds m_i^2 \over \ds m^2_b}, i=\tau,c$. The $W^\pm$ and $H^\pm$
contributions, W(x,y) and H(x,y), are:
$$\eqalign{ W(x,y)=&(2-x-y)(-1-\xi_{\tau}-\xi_c+x+y)\cr
H(x,y)=&y(1+\xi_{\tau}-\xi_c-y)+x(1+\xi_c-\xi_{\tau}-x)\cr
& -(2-x-y)(-1-\xi_{\tau}-\xi_c+x+y)}$$
and the integration limits for $y$ are given by $$y^{\pm}={2 E^{max,min}_{\tau}
\over
m_b},$$ where $E_{\tau}$ is the energy of $\tau$ in the b rest frame.
This result differs from the one presented in Ref. [\isi], because
in our case there is only a new right-handed term, then there
are no interference terms. The numerical result for
the SM contribution to the decay width is ($m_b=4.8~GeV, m_c=1.5~GeV$):
$$ \Gamma_{SM}= |V_{cb}|^2 5.03\times 10^{-9}~MeV, $$
which gives $BR_{SM}\simeq 2.5\%$ for $\mid V_{cb}\mid=0.046\pm 0.005$ [\pdat].

Then, by using the experimental value for $b\to c\tau\nu$
 and the SM result in Eq.(3.14), we obtain the following bound:

$$ \lambda_{bc}\lambda_{\tau\nu} <9.6\times 10^{-2} $$

\chapter{ \bf Predictions from the model}
In this section we use the bounds on $\lambda_{ij}$ obtained in the previous
section to make some predictions of the model.
\section{The decay $c\to u+\gamma$}

One process which could receive an important contribution from the charged
Higgs, is the decay $ c \to u+\gamma$. Here we could expect
a similar behavior as in top quark [\txhx], where the contribution
from the charged and neutral Higgs can give $BR(t\to c\gamma)<10^{-6}$,
which is 6 orders of magnitude larger than the SM prediction.
The Feynman graphs are shown in fig. 1.
The SM contribution will be omited, since the resulting BR
is negligible
( $O(10^{-15}$)) [\egli].
Using the methods of Ref. [\wei] to evaluate the FCNC decays width of
$q_i\to q_j+\gamma$:
$$ \Gamma(c\to u+\gamma)=
{\alpha {G^2_F} {m^5_c} \lambda^2_{bc}\lambda^2_{bu} \over 128\pi^4}
[I_1-{I_2\over 3}]^2, \eqno\eq $$
where the functions that result from the evaluation of the loop-integrals are
\foot{The form of these functions agree with the ones presented in the
literature [\wei], provided one takes the appropiate limit.}:
$$I_1= (1-6z+3z^2+2z^3-6z^2logz)/(1-z)^4,$$
$$I_2=(2+3z-6z^2+z^3+6zlogz)/(1-z)^4,$$\nextline
with  $z={\ds m^2_b\over \ds m^2_H}$.
We have using the approximation $m^2_b>>m^2_c>>m^2_u$.

The branching ratio can be evaluated as follows,
$$ BR(c\to u+\gamma)=
    { \Gamma(c\to u+\gamma) \over \Gamma_{SM}(c\to s+l\nu_l) }
                     BR^{exp}(c\to s+l\nu_l). \eqno\eq $$
Using the SM result for the semileptonic decay of a heavy quark Eq.(3.15), we
obtain:
$${ \Gamma(c\to u+\gamma) \over \Gamma_{SM}(c\to s\ell\nu_{\ell}) } =
      {3\alpha \over 2\pi} [I_1-I_2/3]^2 {\lambda^2_{bc}\lambda^2_{bu}\over
\mid V_{cs}\mid ^2}. $$
Then, using the experimental result for the BR of the
semileptonic decay, namely $BR(c\to s+\ell\nu)=(17.2\pm1.9)\%$
[\pdat] and taking the bound $\lambda_{bc}\lambda_{bu}<6.24\times 10^{-3}$, we
obtain:
$ BR(c\to u+\gamma) < 1.8 \times 10^{-11}$, for $\mid V_{cs}\mid\simeq 1$, and
$z=1/300$, which is well above the SM prediction, but still far from
the experimental limits. Thus, it is unlikely that this result can be tested
at the proposed tau-charm factory [\taus], which is expected to
produce only  $O(10^8)$ c-pairs.

If the QCD corrections, as estimated in [\egli], are included, the BR increases
one order of magnitud, but this value is still far from the experimental reach
\foot{ The inclusion of the neutral Higgs bosons in the loops could enhance
this result, but in that case the analysis depends on more free parameters.}.

\section{ Second class-currents in tau decays}

Another application of the bounds obtained before is on the rate for the
second-class
decay $\tau \to \nu_{\tau}+ \eta \pi$, which has been bounded
experimentaly, namely $BR< 3.4\times 10^{-4}$ [\tuso],
which is still above the SM prediction due to isospin breaking [\ymau] $\simeq
10^{-6}$.
Charged scalar can mediate this process and give a genuine contribution to
second class-current [\yan].

The contribution of the charged Higgs to this decay could be enhanced by the
scalar resonance $a^-_0(980)$. The corresponding amplitude can be written as
 follows,

$$ M= {4G_F V^*_{ud}\lambda_{ud}\lambda_{\tau\nu} \over \sqrt{2} }
      \bar{u_\nu} ( 1-\gamma_5) u_{\tau}
     <\eta\pi | \bar{u} d|0>. \eqno\eq $$
The hadronic matrix element is written as:
$$ <\eta\pi| u\bar d|0> = {S_{a_0}g_{\eta\pi a_0} \over
m^2_{a_0}-q^2-i\Gamma_{a_0}m_{a_0} } ,\eqno\eq $$
where $q=(P_\tau-p_{\nu})$, and the scalar coupling of the $a_0$ is defined
as \nextline $ <a_0| \bar u d|0> = S_{a_0}   $.
 This constant can
be written in terms of the vector coupling of the $a_0$
($ <a_0| \bar u \gamma_\mu  d|0> = if_{a_0} p_\mu e^{-ipx} $), namely [\yan]:
$S_{a_0}=f_{a_0}m^2_{a_0}/(m_d-m_u)\simeq (0.55~GeV)^2$ where $f_{a_0}\simeq
2~MeV$ [\yan]. The $a_0\eta\pi$ coupling
constant is extracted from the total width of the $a^-_0$, giving
$g_{a^-_0\eta\pi}=(2.02\pm0.13)~GeV$.

Substituting the previous relations in the squared amplitude and
including the phase-space factor, one finds:
$$\Gamma(\tau\to\nu_{\tau}+\eta\pi)= C_0 S^2_{a_0} \lambda^2_{ud}
\lambda^2_{\tau\nu_{\tau}} \eqno\eq$$
where $C_0=(1.15\pm 0.22)\times 10^{-11}~GeV^{-3}$.
In order to obtain a numerical result, we shall use the bound obtained
for $\lambda_{ud}\lambda_{\tau\nu_{\tau}}$ Eq.(3.11), which was extracted from
$\tau\to \nu_{\tau}\pi$ and leptonic tau decay.
Thus we find:
$BR(\tau\to \nu+\eta\pi)< 4\times10^{-5}$, which is consistent with the present
experimental bound~[\pdat]. If the $\lambda_{ij}`s$ saturate the bound on their
values, the BR would be above the SM prediction, and this could be testable at
the proposed tau
-charm factories~[\taus].

\section{Neutron Beta decay}

  Another interesting point to be considered within our model is
neutron beta decay. Within the standard model, the neutron
lifetime can be predicted [\cast], in terms of the ratio of axial and
vector form factors, $\lambda' = g_A/f_V=-1.2573 \pm 0.0028$ [\pdat], and
the quark mixing angle $V_{ud}=0.9744 \pm 0.0010$ [\pdat].
The value obtained for the neutron lifetime,
$\tau_n = (900 \pm 4)~sec$,
is two standard deviations above the measured
\foot{ It is important to mention that both
values for $\tau_n$ agree if $\mid \lambda'\mid$ increases up to 1.266 .}
value
$\tau_n = (889.1 \pm 2.1)~sec$ [\pdat].
As it will be shown below, the charged Higgs contribution of our model
could provide a solution to this discrepancy.

   Following reference [\habe] we write an effective Hamiltonian for
neutron beta decay, including $W^\pm$ and $H^\pm$ contributions, as follows:
$$\eqalign{H_{eff}=&{G_F \over \sqrt{2}}V_{ud}\big[\bar
u_p\gamma_{\mu}(f_V+g_A\gamma_5)u_n\bar u_e\gamma_{\mu}(1-\gamma_5)u_{\nu_e}
\cr
&+{ \lambda_{ud} \over V_{ud}}\lambda_{e\nu_e}\bar u_p(f_S+g_P\gamma_5)u_n\bar
u_e(1-\gamma_5)u_{\nu_e}\big]}, \eqno\eq$$
where the scalar and pseudoscalar form factors, $f_S$ and $g_P$, arise
from the Higgs mediated interaction. We can write the form factor $f_S$
in terms of charged Higgs parameters as follows:

$$ \lambda^{''} \equiv f_S/f_V= {\lambda_{ud}\lambda_{e\nu}(m_n-m_p)
              \over V_{ud}(m_d-m_u) }.$$
We do not write the corresponding expression for the $g_P$ form
factor because it gives a negligible contributions to the neutron decay rate.

   Thus, the total decay rate can be written as follows:
$$
\eqalign{\tau_n =&{ 1 \over  \Gamma(n\to pe^-\nu_e)}\cr
=&{ 2\pi^3(1-\Delta_{\beta}+\Delta_{\mu})}\over { G^2_Fm^5_e\mid V_{ud}\mid ^2
(1+3\lambda'^2+\lambda''^2)f_1}}, \eqno\eq$$
where $f_1 = 1.71645 \pm 0.00015$  is the phase space factor including
{\em outer} [\wilk] e.m. radiative corrections. 
The {\em inner} e.m. radiative corrections are included in
$\Delta_{\beta} - \Delta_{\mu} \simeq 2.34 \%$.
A comparison of (eq. 4.7) with $\tau_n^{exp}$
gives $f_S/f_V \simeq 0.2477$ or equivalently,
$\lambda_{ud}\lambda_{e\nu}\simeq0.8024$, which seems too large as compared
with the values derived in the previous sections.

\chapter{\bf Conclusions}

In summary, we have studied several low-energy
proccess to constraint the parameters of our version of the
two-Higgs doublet model. Table 1 summarizes the bounds on the effective
$H^{\pm}f_i\overline f_j$ couplings defined in Eq.(3.1). To our knowledge this
is the first study of generalized Yukawa couplings of the charged Higgs boson.

Through the use of these constraints, it has
been possible to study several predictions of the model.
In particular,
we find that the model predicts BR($c\to u+\gamma)<10^{-11}$, which is much
larger than the SM predictions [\egli].
Unfortunately, this result can not be tested at the proposed tau-charm factory
[\taus]. The prediction of large FCNC for the u-type
quarks is clearly different from the SM picture.

Similarly, we find that the contribution from a charged Higgs
boson to the second-class current decay of the tau give the bound $BR\leq
10^{-5}$,
which is consistent with the present experimental bound. If the
$\lambda_{ij}`s$ saturate the bound on their values, the BR would be above the
SM prediction, and this could be testable at the proposed tau-charm factories
[\taus].

Finally, the neutron lifetime would require a too large contribution from the
charged Higgs, which seems inconsistent with the bounds obtained in section 3.

As we mentioned in the text, by studying the Yukawa couplings
of the charged Higgs, it may be possible to test the models that attempt to
explain the fermion mass spectrum of the SM. Even if some model can explain the
spectrum, it would require further predictions in order to be accepted as a
viable theory
 of masses. Our proposal is to search for a charged Higgs and study its Yukawa
couplings, to see if it has the form predicted in those models that use an
ansatz for the fermionic mass matrices [14], which give definite predictions
for the Yukawa interacti
ons of the Higgs bosons.
\foot{ However, at present such comparison is not possible without knowing the
charged Higgs mass, which appears in our definition of $\lambda `s$.}.
\vskip 4pc

\ack{  We acknowledge the finantial support from
CONACYT(ME- \break
XICO). }

\endpage

\refout
\endpage
\FIG\feync{ Feynman graphs for the contribution of charged Higgs to
the 1-loop amplitude for $c\to u+\gamma$. }

\figout
\TABLE\ta{ Summary of bound the $\lambda_{ij}$'s obtained from low energy data
as explained in the text. }

\tabout

\end